\documentclass[twocolumn,
secnumarabic,
amssymb,
nobibnotes,
aps,
prc,
superscriptaddress,
longbibliography]{revtex4-1}

\usepackage{graphicx}
\usepackage{longtable}
\usepackage[colorlinks,
           linkcolor=blue,
            anchorcolor=black,
            citecolor=blue
            ]{hyperref}
\usepackage{amsmath}
\usepackage[normalem]{ulem}
\usepackage[dvips]{color}
\usepackage{multirow}
\usepackage{appendix}
\usepackage{CJK}
\usepackage{lineno}
\begin{document}

\title{\textbf{Accessing the in-medium effects on nucleon-nucleon elastic cross section with collective flows and nuclear stopping}}
\author{Pengcheng Li}
\affiliation{School of Nuclear Science and Technology, Lanzhou University, Lanzhou 730000, China}
\affiliation{School of Science, Huzhou University, Huzhou 313000, China}

\author{Yongjia Wang}
\affiliation{School of Science, Huzhou University, Huzhou 313000, China}

\author{Qingfeng Li}
\thanks{Corresponding author: liqf@zjhu.edu.cn}
\affiliation{School of Science, Huzhou University, Huzhou 313000, China}
\affiliation{Institute of Modern Physics, Chinese Academy of Sciences, Lanzhou 730000, China}

\author{Hongfei Zhang}
\affiliation{School of Physics, Xi'an Jiaotong University, Xi'an 710049, China}
\date{\today}

\begin{abstract}
A systematic study of the in-medium correction factor ($F$) on nucleon-nucleon elastic cross section is performed within the Ultra-relativistic Quantum Molecular Dynamics (UrQMD) model.
The effects of the beam energy dependence of $F$ on the directed, elliptic flow and nuclear stopping in $^{197}$Au+$^{197}$Au collisions with energy ranging from $0.09$  to $0.8A$ GeV are explored.
It is found that the directed, elliptic flow and nuclear stopping at relatively low energies are very sensitive to $F$, and the sensitivity gradually weakens with increasing beam energy.
The beam energy dependent in-medium correction factor $F$ is deduced from the comparison of the excitation functions of the directed, elliptic flow and nuclear stopping between the calculated results and the FOPI experimental data.

\end{abstract}

%

\maketitle

\section{Introduction}\label{section1}
The main goal of heavy-ion physics at intermediate energies is to explore the properties of the hot and dense strongly interacting nuclear matter.
Comparing experimental data from terrestrial laboratories to theoretical calculations is one of the commonly used method to explore the fundamental properties of nuclear matter under a wide range of densities, temperatures, and isospin asymmetries \cite{Danielewicz:2002pu,Li:2008gp,Xu:2019hqg,Colonna:2020euy}. Boltzmann-Vlasov-type (usually referred to BUU-type) and molecular dynamics-type (usually referred to QMD-type) models are two of the most popular theoretical models for simulating heavy-ion collisions (HICs) at intermediate energies.
The in-medium nucleon-nucleon elastic cross section ($NNECS$) is one of the important ingredients of both models, it has been widely investigated in recent decades \cite{Xu:2016lue,Zhang:2017esm,Ono:2019ndq,Colonna:2021xuh}.

The $NNECS$ in free space $\sigma^{\text{free}}_{\rm{el}}$ can be directly measured by experiments.
However, the information of the in-medium $NNECS$ ($\sigma^{\rm{\text{in-med}}}_{\rm{el}}$) usually is constrained by theoretical assumptions.
These theoretical calculations include, but not limited to, the Dirac-Brueckner approach with Bonn potential \cite{Li:1993rwa,Li:1993ef}, the Dirac-Brueckner-Hartree-Fock approach with realistic nucleon-nucleon potential \cite{Sammarruca:2005tk}, the relativistic Brueckner-Hartree-Fock model \cite{Zhang:2007zzs,Zhang:2010jf}, the closed time-path Green's function approach \cite{Mao:1994zza}.
It is shown definitely that the $\sigma^{\rm{\text{in-med}}}_{\rm{el}}$ is modified by the nuclear medium.
However, the degree of this modification is still far from being solved thoroughly.

In most of theoretical models that used to simulate HICs at intermediate energies, the parameterized in-medium correction factor on $NNECS$ is commonly used for simplicity.
In general, this correction factor $F=\sigma^{\rm{\text{in-med}}}_{\rm{el}}/\sigma^{\text{free}}_{\rm{el}}$ is density- and/or momentum-, as well as isospin-dependent \cite{Li:2005iba,Zhang:2007gd,appb33452002,Li:2011zzp,Coupland:2011px,prc72024611,Su:2016adl}.
Many model simulations have demonstrated that various phenomena in HICs are sensitive to $\sigma^{\rm{\text{in-med}}}_{\rm{el}}$, thus the final state observables of HICs can be selected, such as the particle yields, collective flows and the nuclear stopping (energy dissipation), to extract the information of $\sigma^{\rm{\text{in-med}}}_{\rm{el}}$
\cite{Li:2011zzp,Wang:2013wca,Li:2018wpv,Li:2005iba,Zhang:2007gd,Su:2016adl,Chen:2021cpc,Lopez:2014dga,Basrak:2016cbo,Henri:2020ezr,Zhang:2020qqs,appb33452002,Coupland:2011px,prc72024611,Wang:2020xgk}.
In Ref.\cite{Cai:1998iv}, a phenomenological formula for the in-medium $NNECS$ which depends on both density and beam energy was proposed, and found that the in-medium effect gradually weakens with increasing beam energy.
By studying the nuclear stopping in central collisions in the Fermi-energy domain, it is also found a reduction on the in-medium effect, e.g., $F$ is about 0.2 at $E_{lab}=0.035A$ GeV and 0.5 at $0.1A$ GeV \cite{Lopez:2014dga}, respectively.
Usually, the beam energy dependence of this reduction factor used in transport models is partly reflected in other physical quantities, such as in the density, momentum and/or simply in $\sigma^{\text{free}}_{\rm{el}}$.

In our previous work, about twenty years ago, the in-medium $NNECS$ was studied based on the extended quantum hadrodynamics model in which the interaction between nucleons is described by exchanges of $\sigma$, $\omega$, $\pi$, $\rho$, and $\delta$ mesons \cite{Li:2000sha,Li:2003vd} .
Several years later, the density-, momentum-, and isospin-dependent in-medium correction factor on $NNECS$ was introduced into the ultra-relativistic quantum molecular dynamics (UrQMD) model \cite{ppnp41255,JPG251859} in Ref. \cite{Li:2006ez}, and the transverse flow as a function of rapidity, the momentum quadrupole as a function of momentum, and the ratio of halfwidths of the transverse to that of longitudinal
rapidity distribution were found to be sensitive to the in-medium correction factor. Later on, a reduction on $\sigma^{\rm{\text{in-med}}}_{\rm{el}}$ compared to $\sigma^{\text{free}}_{\rm{el}}$ was deduced from the comparison of the nuclear stopping data at SIS energies\cite{Yuan:2010ad}, of the collective flows data at INDRA energies \cite{Li:2011zzp,Li:2018wpv,Wang:2013wca}. However, a systematical deduction on $\sigma^{\rm{\text{in-med}}}_{\rm{el}}$ from both the nuclear stopping and collective flows data over INDRA and SIS energies is still missing.

The purpose of this work is to explore the beam energy dependence of the in-medium $NNECS$ over a wide range of beam energy by using both the collective flows and nuclear stopping data. The simulations are performed on $^{197}$Au+$^{197}$Au collisions by the UrQMD model. This paper is organized as follows: in Sec. \ref{sec2}, the UrQMD model and the observables will be briefly recalled. In Sec. \ref{sec3} the energy dependence of the in-medium $NNECS$ and its influence on collective flows and nuclear stopping in HICs at SIS energies are shown. Finally the conclusions and outlooks are presented in Sec. \ref{sec4}.

\section{the UrQMD model and observables} \label{sec2}
In the UrQMD model, each nucleon is represented by a coherent state of a Gaussian wave packet. And the coordinate $\textbf{r}_i$ and momentum $\textbf{p}_i$ of $i$-th nucleon are propagated according to Hamilton's equation of motion \cite{Li:2011zzp}
$\dot{\textbf{r}}_{i}=\frac{\partial  \langle H  \rangle}{\partial\textbf{ p}_{i}}$,
$\dot{\textbf{p}}_{i}=-\frac{\partial  \langle H \rangle}{\partial \textbf{r}_{i}}$.
Here, {$ \langle H \rangle$} is the total Hamiltonian function of the system, comprising the kinetic energy and the potential energy.
For studying HICs at intermediate energies, the following density- and momentum-dependent potential form is frequently used in QMD-like models \cite{Aichelin:1991xy,Hartnack:1997ez},
\begin{equation}\label{eq2}
V=\alpha \cdot (\frac{\rho}{\rho_0})+\beta \cdot (\frac{\rho}{\rho_0})^{\gamma} + t_{md} \ln^2[1+a_{md}(\textbf{p}_{i}-\textbf{p}_{j})^2]\frac{\rho}{\rho_0}.
\end{equation}
Where $\alpha$=-393 MeV, $\beta$=320 MeV, $\gamma$=1.14, $t_{md}$=1.57 MeV, and $a_{md}=500$ $c^{2}$/GeV$^{2}$ are adopted in this work, which yields a soft and momentum-dependent equation of state with the incompressibility $K_{0}=200~\rm{MeV}$. It has been checked that varying $K_0$ within its presently accepted constraint (e.g., $K_0=200\sim280$ MeV) \cite{Danielewicz:2002pu,Xu:2019hqg,Wang:2018hsw,LeFevre:2015paj,Li:1999bh,Li:2021ikk,Xu:2021aij} will not affect significantly the results of present work discussed below.
It is known that the Pauli blocking plays a vital role in theoretical study of HICs in the low to intermediate energies \cite{Zhang:2017esm,Li:2011zzp,Chen:2021cpc}.
But different Pauli blocking algorithms are used in different transport models, and the effectiveness of these Pauli blocking algorithms differ substantially among the different transport model codes \cite{Zhang:2017esm}.
With the same treatments on mean fields and $NN$ cross sections for the same combination system, the uncertainty of the $v_{1}$ slope from different transport codes, where different Pauli blocking algorithms exist, is about 13\% (30\%) at the beam energy 0.4 (0.1)$A$ GeV \cite{Xu:2016lue}.
By using three typical Pauli blocking algorithms (PB-Wigner (which is used in the current version of the UrQMD code), PB-Husimi, and PB-HSP), which are adopted in the different QMD-type models, the effects of different Pauli blocking algorithms on the excitation function of stopping power in HICs are analyzed.
It is found that the uncertainties are less than 5\% at beam energies below $0.3A$ GeV \cite{Chen:2021cpc}.
For simply, the Pauli blocking algorithm used in this work is set in the same way as our previous study \cite{Li:2011zzp}, but the Pauli blocking algorithms used in transport models and its effect on observables certainly deserves further studies.

In the UrQMD model, the $\sigma_{\text{el}}^{\rm{\text{in-med}}}$ is treated to be factorized as the product of a medium correction factor $\mathcal{F}(\rho,p)$ and the cross sections in free space for which the experimental data are available,
\begin{equation}
\sigma_{\text{el}}^{\rm{\text{in-med}}}=\mathcal{F}(\rho,p)\cdot\sigma^{\text{free}}_{\rm{el}}
\end{equation}
with
\begin{equation}
\mathcal{F}(\rho,p)=\left\{
\begin{array}{l}
f_0, \hspace{2.8cm} p_{NN}>1 {\rm GeV}/c, \\
\frac{\lambda+(1-\lambda)e^{-\frac{\rho}{\zeta\rho_{0}}} -f_{0}}{1+(p_{NN}/p_{0})^\kappa}+f_{0}, \hspace{0.5cm} p_{NN} \leq 1 {\rm GeV}/c.
\end{array}
\right.
\label{fdpup}
\end{equation}
Where $f_{0}$, $\lambda$, $\zeta$, $\kappa$ are the parameters and $p_{NN}$ is the momentum in the two-nucleon center-of-mass frame.
In this work, the FU3FP2 and FU3FP4 parametrization of $NNECS$ are adopted as we did in our previous works \cite{Li:2011zzp,Wang:2020vwb,Wang:2013wca,Wang:2014rva}, and the parameter sets are listed in Tab. \ref{tabfufp}. Here, $f_{0}>1 (<1)$ implies a possibly enhanced (reduced) in-medium effect on $NNECS$ for $p_{NN}>1$ GeV/$c$.
\begin{table}[h]
\centering
\caption{\label{tabfufp} The parameter sets FU3FP2 and FU3FP4 used for describing the density and momentum dependence of $\mathcal{F}(\rho,p)$.}
\setlength{\tabcolsep}{2mm}
\begin{tabular}{c|ccccc}
\hline\hline
             & $\lambda$     & $\zeta$     & $f_{0}$     &$p_{0}$     & $\kappa$   \\ \hline
FU3FP2       & $1/6$         & $1/3$       & $1$         &$0.225$     & $3$        \\
FU3FP4       & $1/6$         & $1/3$       & $1$         &$0.3$       & $8$        \\ \hline\hline
\end{tabular}
\end{table}

Based on our previous investigations, a stronger momentum reduction parametrization (FU3FP1) is used for reproducing experimental data at the Fermi energy region while the weaker FU3FP4 is used to extract the nuclear incompressibility and the density-dependent symmetry energy from the elliptic flow at higher energies.
To better reflect the beam energy dependence of the in-medium correction on $NNECS$ over a wide beam energy region, the factor $\tanh(E_{lab}/\varepsilon)$ is introduced into Eq.\ref{fdpup} so that
\begin{equation}
\mathcal{F}(\rho,p)=\left\{
\begin{array}{l}
f_0, \hspace{4.3cm} p_{NN}>1 {\rm GeV}/c, \\
\tanh(\frac{E_{lab}}{\varepsilon})[\frac{\lambda+(1-\lambda)e^{-\frac{\rho}{\zeta\rho_{0}}} -f_{0}}{1+(p_{NN}/p_{0})^\kappa}+f_{0}], \hspace{0.5cm}
p_{NN} \leq 1 {\rm GeV}/c.
\end{array}
\right.
\end{equation}
Here, the parameter $\varepsilon$ is set to be 0.2.
This form is inspired by the in-medium correction factors used in pBUU model \cite{appb33452002}, and the momentum dependence of $\mathcal{F}(\rho,p)$ \cite{Wang:2020vwb}.
The extended in-medium correction factors are labeled as $\varepsilon$FU3FP2 and $\varepsilon$FU3FP4.
In addition, on the basis of our previous work \cite{Li:2018wpv},
the fixed equivalent in-medium correction factors $F=\sigma_{\rm{el}}^{\rm{in-med}}/\sigma^{\rm{free}}_{\rm{el}}$ at different energies are further considered in this work.

In general, the directed, elliptic flow, and the nuclear stopping in HICs at intermediate energies are strongly related to the two-body scatters, and have been widely studied \cite{Li:2008gp,Li:2011zzp,Basrak:2016cbo}.
The directed ($v_{1}=\left\langle\frac{p_{x}}{\sqrt{p_{x}^{2}+p_{y}^{2}}}\right\rangle$) and elliptic ($v_{2}=\left\langle\frac{p_{x}^{2}-p_{y}^{2}}{p_{x}^{2}+p_{y}^{2}}\right\rangle$) flow can be deduced from the Fourier expansion of the azimuthal distribution of detected particles \cite{Reisdorf:1997fx}.
The angle brackets indicate an average over all considered particles from all events.
The nuclear stopping governs the amount of dissipated energy, i.e. the efficiency of converting the beam energy in the longitudinal direction into the transverse direction. Serval different quantities of nuclear stopping have been used and investigated \cite{Liu:2001uc,FOPI:2004orn,INDRA:2010pbz,FOPI:2010xrt,Shi:2021far}.
In order to compare with FOPI data, we mainly focus on $varxz$ in the present work.
It is defined as the ratio of the variances of particle rapidity distribution along the transverse $\Gamma^{2}(y_{x})$ to those of the longitudinal $\Gamma^{2}(y_{z})$ rapidity distribution\cite{FOPI:2006ifg}, which read as $\textit{varxz} = \frac{\Gamma^{2}(y_{x})}{\Gamma^{2}(y_{z})}$.
Here, $varxz=1$ is corresponding to an isotropic thermal source, the energy distribution is isotropical.
While $varxz>1$, the energy is preferentially distributed in the transverse plane of the reaction.
Once $varxz<1$, the energy is preferentially distributed along the beam direction \cite{FOPI:2004orn,INDRA:2010pbz,Henri:2020ezr}.

\section{results}\label{sec3}
\subsection{Collective flows}\label{sec31}
\begin{figure}[t]
\begin{center}
\includegraphics*[scale=0.32]{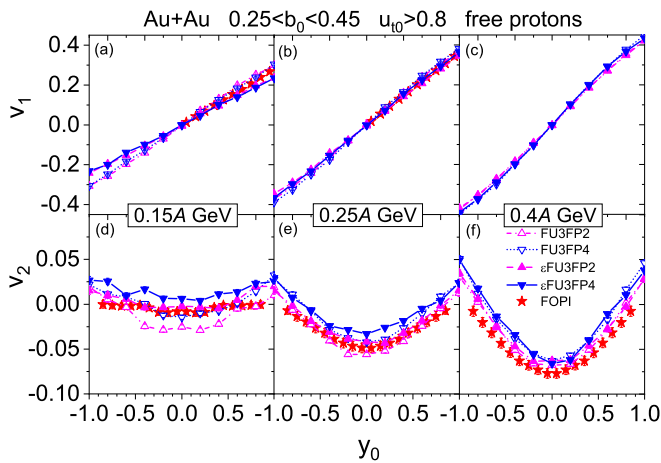}
\caption{(Color online) Reduced rapidity $y_{0}$ distributions of the directed flow $v_{1}$ (top panels) and elliptic flow $v_{2}$ (bottom panels) of free protons from semi-central Au+Au collisions, with transverse 4-velocities cut $u_{t0}>0.8$. The calculated results with four in-medium correction factors are presented by different symbols as indicated, and the FOPI data are taken from Ref. \cite{FOPI:2011aa}.}
\label{v12y0}
\end{center}
\end{figure}

\begin{figure}[t]
\begin{center}
\includegraphics*[scale=0.35]{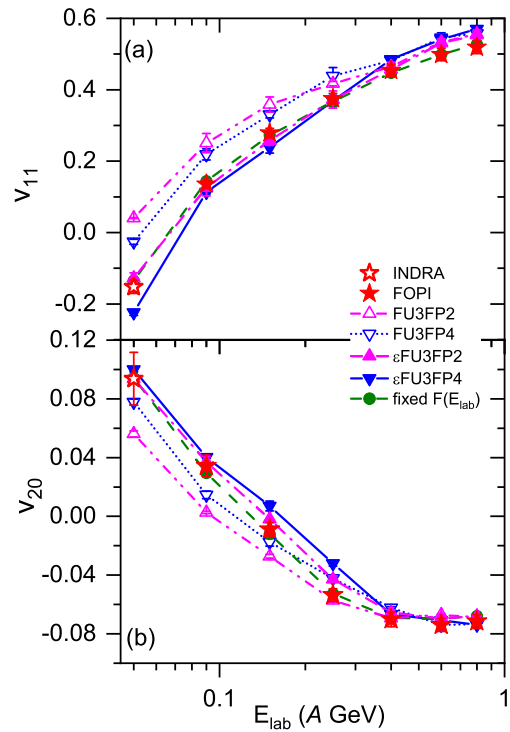}
\caption{(Color online) Beam energy dependence of directed flow slope $v_{11}$ (top panel) and elliptic flow at mid-rapidity $v_{20}$ (bottom panel) from semi-central Au+Au collisions. The $v_{11}$ and $v_{20}$ are obtained with assuming
$v_{1}(y_{0})=v_{11}\cdot y_{0}+v_{13}\cdot y_{0}^{3}+c$ and $v_{0}(y_{0})=v_{20}+v_{22}\cdot y_{0}^{2}+v_{24}\cdot y_{0}^{4}$ in the range of $|y_{0}|\leqslant0.6$. The FOPI and INDRA are taken from Ref. \cite{FOPI:2011aa,LeFevre:2016vpp,Andronic:2006ra,Russotto:2013fza}, respectively. Shown as open stars are the results from INDRA for Z=1 particles, while solid stars are the results from FOPI for free protons and $u_{t0}>0.8$. The different curves (symbols) are calculated with various in-medium correction factors (see text for details).}
\label{v11v20}
\end{center}
\end{figure}

Fig. \ref{v12y0} shows the directed (top panels) and elliptic (bottom panels) flow of free protons in semi-central Au+Au collisions at $E_{lab}=$ 0.15, 0.25, $0.4A$ GeV, which are simulated with different in-medium correction factors on $NNECS$.
The intervals of the reduced impact parameter $b_{0}$ and the scaled transverse velocity $u_{t0}$ are
chosen to be the same as in the FOPI analysis \cite{FOPI:2011aa}, i.e., $0.25<b_{0}<0.45$ and $u_{t0}>0.8$, respectively.
These quantities are defined as $y_{0}=y_{z}/y_{pro}$ with $y_{pro}$ being the projectile rapidity in the center-of-mass system, $b_{0}=b/b_{max}$ with $b_{max} = 1.15 (A _{P}^{1/3} + A_{T}^{1/3})\mbox{ fm}$, $u_{t0}\equiv u_{t}/u_{pro}$ with $u_{t}=\beta_{t}\gamma $ the transverse component of the four-velocity and $u_{pro}$ is the velocity of the incident projectile in the center-of-mass system \cite{FOPI:2011aa}.
The results from calculations with FU3FP2, FU3FP4, $\varepsilon$FU3FP2, and $\varepsilon$FU3FP4 are represented by open up triangles, open down triangles, solid up triangles, and solid down triangles,
respectively.
A good agreement between the FOPI data and the model calculations in the whole rapidity range can be found.
At $0.15A$ GeV, $v_{1}$ calculated with FU3FP2 and FU3FP4, i.e., without beam energy dependence of $\sigma^{\text{in-med}}_{\rm{el}}$, are slightly larger than that with $\varepsilon$FU3FP2 and $\varepsilon$FU3FP4, but these differences are small at $0.25$ and $0.4A$ GeV.
For $v_{2}$, the gaps between the calculations with $\varepsilon$FUFP and FUFP sets are visibly and become gradually smaller with increasing beam energy, and finally almost unrecognizable at $0.4A$ GeV. These results imply that the collective flows are sensitive to the beam energy dependence of the in-medium $NNECS$, especially at relatively low energies.

In order to quantitatively show the influence of the beam energy dependence of the in-medium $NNECS$ on the collective flows, the $v_{1}$ slope and $v_{2}$ at mid-rapidity for free protons are calculated
and compared to the experimental data \cite{FOPI:2011aa}, and displayed in Fig. \ref{v11v20}.
As expected both $v_{11}$ and $v_{20}$ calculated with FUFP and $\varepsilon$FUFP sets are well separated at low beam energies, and the difference vanishes at high energies. 
It is known in our previous work \cite{Li:2018wpv,Wang:2013wca,Li:2011zzp,Wang:2018hsw,Wang:2020vwb}, that a stronger reduction parametrization (e.g., FU3FP1) in $\sigma_{\rm{el}}^{\rm{in-med}}$ is required to reproduce flow and stopping data at the Fermi energy, while a relatively weak reduction parametrization (e.g., FU3FP4) is favoured at $E_{lab}\gtrsim0.25A$ GeV. From Fig. \ref{v11v20}, it can be seen that both $v_{11}$ and $v_{20}$ from 0.04A GeV to 1.0A GeV can be well reproduced with $\varepsilon$FU3FP2.

\begin{figure}[t]
\begin{center}
\includegraphics*[scale=0.37]{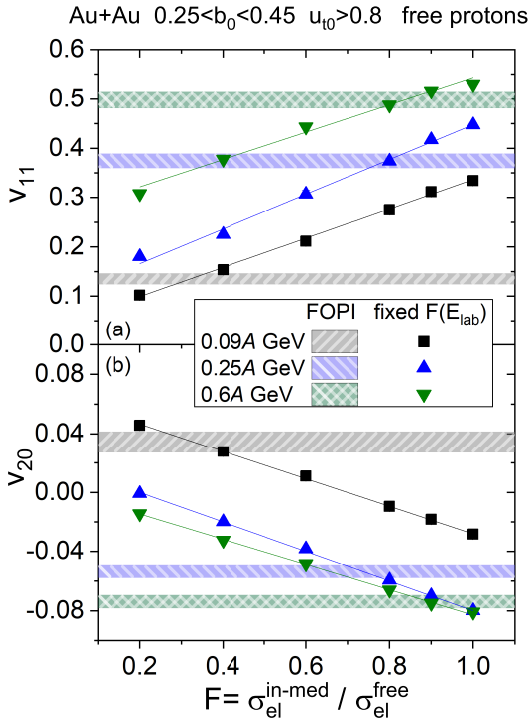}
\caption{(Color online) The $v_{11}$ (top panel) and $v_{20}$ (bottom panel) for free protons from semi-central Au+Au collisions with $u_{t0}>0.8$ are plotted as a function of the in-medium correction factor. 
The solid curves are linear fits to the calculated results, and the shaded bands are the corresponding experimental data \cite{FOPI:2011aa,LeFevre:2016vpp}.}
\label{varxz-e}
\end{center}
\end{figure}

\begin{figure}[htbp]
\begin{center}
\includegraphics*[scale=0.35]{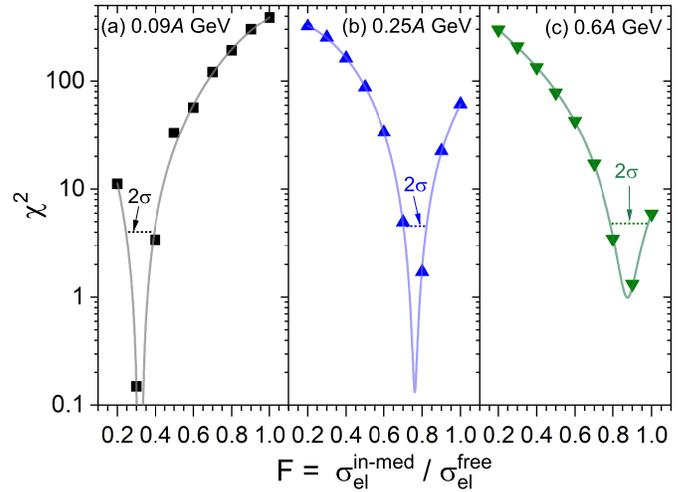}
\caption{(Color online) The $\chi^{2}$ as a function of the in-medium correction factor of $NN$ elastic collisions. The smooth solid curves are the quadratic fit to the total $\chi^{2}$ obtained from free protons $v_{11}$ and $v_{20}$, and the short dot lines are used to determine the error of $F$ within a 2-$\sigma$ uncertainty.}
\label{chi2}
\end{center}
\end{figure}

It is of interest to quantitatively understand the in-medium effect on $NNECS$ 
at different beam energies, the same as our previous work \cite{Li:2018wpv},
$v_{11}$ and $v_{20}$ of free protons calculated with the fixed $F$ at various energies are shown in Fig. \ref{varxz-e}.
A fairly well linear relationship between the $v_{11}$ ($v_{20}$) and $F$ can be seen, confirming that the collective flows are indeed sensitive to the in-medium effects.
Consequently, one can constrain the value of $F$ at each beam energy with the chi-square analysis. The $\chi^{2}=\sum_{i}\frac{(X^{\rm{th}}_{i}-X^{\rm{exp}}_{i})^{2}}{\sigma_{i}^{2}}$ is plotted as a function of $F$ in Fig. \ref{chi2}, where $X^{\rm{th}}_{i}$ and $X^{\rm{exp}}_{i}$ is the theoretical and the corresponding experimental values and $\sigma_{i}$ is the theoretical error.
There is also exits a well linear relationship between nuclear stopping and $F$, however, the $F$ constrained from nuclear stopping have a large error, since the precision of nuclear stopping data is much worse than that of collective flows.
The obtained $F$ with a 2-$\sigma$ confidence limit (at 95\% confidence level) based on the comparison of the FOPI experimental data \cite{FOPI:2011aa,LeFevre:2016vpp} on the collective flows of free protons with the UrQMD model calculations are shown in Tab. \ref{tab1}. One can find that this correction factor is beam energy-dependent, and by adopting the extracted $F$ into simulations, the $v_{11}$ and $v_{20}$ data can be described fairly well, as shown in Fig.\ref{v11v20}.

\begin{table}[htbp]
\centering
\caption{\label{tab1} The extracted equivalent energy-dependent in-medium correction factor $F$, with a 2-$\sigma$ confidence limit.
}
\setlength{\tabcolsep}{1.1mm}
\begin{tabular}{c|cccccccc}
\hline\hline
$E_{lab}$ ($A$ GeV)       & $0.09$     & $0.15$     & $0.25$     &$0.4$     & $0.6$      & $0.8$       \\ \hline
fixed $F$    & $0.32$     & $0.57$      & $0.76$     &$0.86$    & $0.87$     & $0.87$     \\
error                     & $\pm0.07$  & $\pm0.06$  & $\pm0.06$  &$\pm0.07$  & $\pm0.10$  & $\pm0.09$  \\ \hline\hline
\end{tabular}
\end{table}

\begin{figure}[t]
\begin{center}
\includegraphics*[scale=0.37]{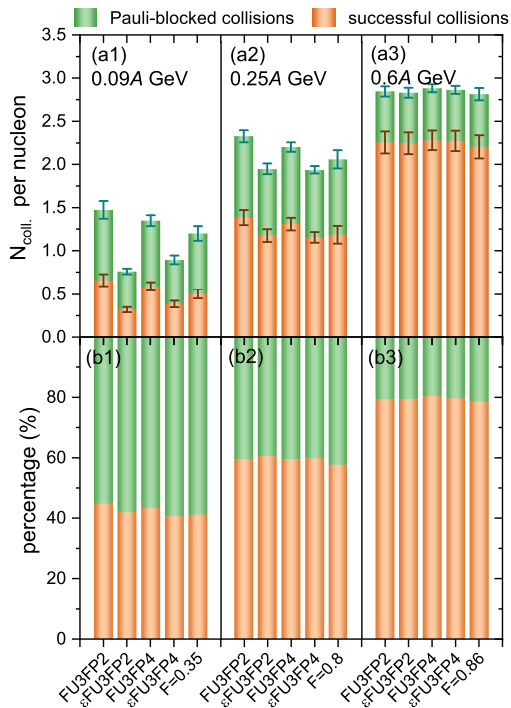}
\caption{(Color online) Top panels (a1-a3): the collisions numbers per nucleon experienced in central Au+Au collisions with different in-medium correction factors. Bottom panels (b1-b3): the percentage of the successful and Pauli-blocked collisions from calculations with different in-medium correction factors (see text for more details).}
\label{dNcoll}
\end{center}
\end{figure}

It is known that nucleon-nucleon collisions will be influenced by the nuclear medium, consequently the number of collisions will be affected by medium effects. The top panels (a1-a3) of Fig. \ref{dNcoll} show the averaged collision number per nucleon experienced in central Au+Au collisions with different in-medium correction factors, the successful collision number and Pauli-blocked number is represented by orange and green bands, respectively. The total collision number from simulations with FU3FP2 and FU3FP4 is larger than that of $\varepsilon$FU3FP2 and $\varepsilon$FU3FP4. For example, the successful collision of FU3FP2 (FU3FP4) is about 42\% (43\%) larger than that of $\varepsilon$FU3FP2 ($\varepsilon$FU3FP4) at $E_{lab}=$ 0.09$A$ GeV, while the collision number among different calculations at $E_{lab}=$ 0.6$A$ GeV are almost the same.
We have checked the relationship between the collision number and the equivalent in-medium correction factor $F$, and found that with increasing the equivalent in-medium correction factor $F$, i.e., decreasing the in-medium effects, the collision number will increase almost linearly. It must be emphasized again that the change of the equivalent fixed $F$ results in a global effect, which gives a correction on all collisions, regardless of the density, momentum, and isospin. And at intermediate energies mentioned in this work, the values of $v_{11}$ ($v_{20}$) will increase (decrease) when the collision number increases. Because nucleons which experience more collisions have larger probability to bounce-off and squeeze-out caused by the presence of the nearby spectator matter. The bottom panels (b1-b3) display the percentage of the successful (Pauli-blocked) collision number to total collisions number. These ratios hardly change when the in-medium correction factors are modified, since the Pauli blocking algorithm is not modified.

\begin{figure}[t]
\begin{center}
\includegraphics*[scale=0.37]{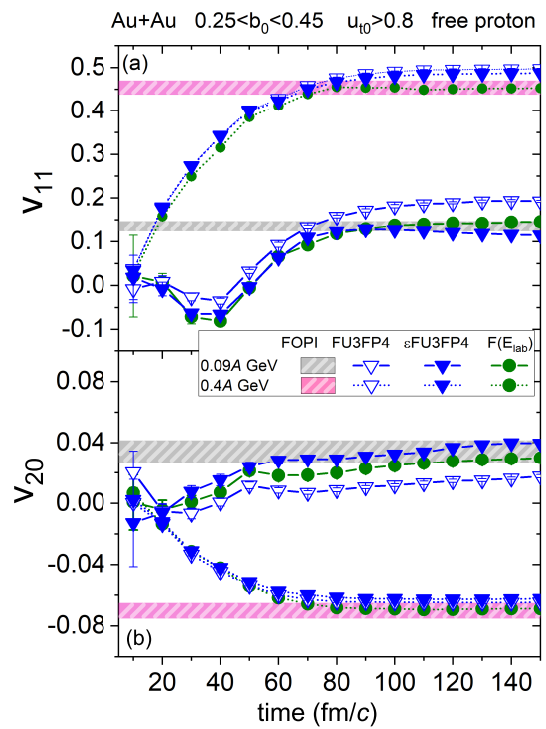}
\caption{(Color online) Time evolution of the slope of direct flow and elliptic flow at midrapidity ($|y_{0}|\leqslant0.6$) for free protons from semi-central Au+Au collisions at $E_{lab}$=0.09 and $0.4A$ GeV. 
The FOPI data \cite{FOPI:2011aa,LeFevre:2016vpp} for both beam energies are indicated by the bands with different color.}
\label{v10v20time}
\end{center}
\end{figure}

The time evolution of the $v_{11}$ and $v_{20}$ for free protons from semi-central Au+Au collisions at $E_{lab}=0.09$ and $0.4A$ GeV are shown in Fig. \ref{v10v20time}.
In panel (a), before $\sim$50 fm$/c$ the value of $v_{11}$ at $0.09A$ GeV is negative, since the net contribution is attractive, which leading to the corresponding negative flow \cite{Guo2012}.
After $\sim$50 fm/c, the final-state interactions still affect the collisions, and the contributions of two-body scatterings start to become stronger than those of the mean-field potentials. While, at $0.4A$ GeV the net contribution is repulsive, leading to the positive flow.
In panel (b), at low energies ($0.09A$ GeV), the negative values for $v_{20}$ can be seen at early time (compressed stage), then the compressed region expands but protons are preferential emitted in plane, reflecting positive values for $v_{20}$ as result of the attractive potentials between the nucleons. At higher energies ($0.4A$ GeV), the strength of the collective expansion will overcome the rotational-like motion, leading to an increase of out-of-plane emission (negative elliptic flow).

\begin{figure}[t]
\begin{center}
\includegraphics*[scale=0.34]{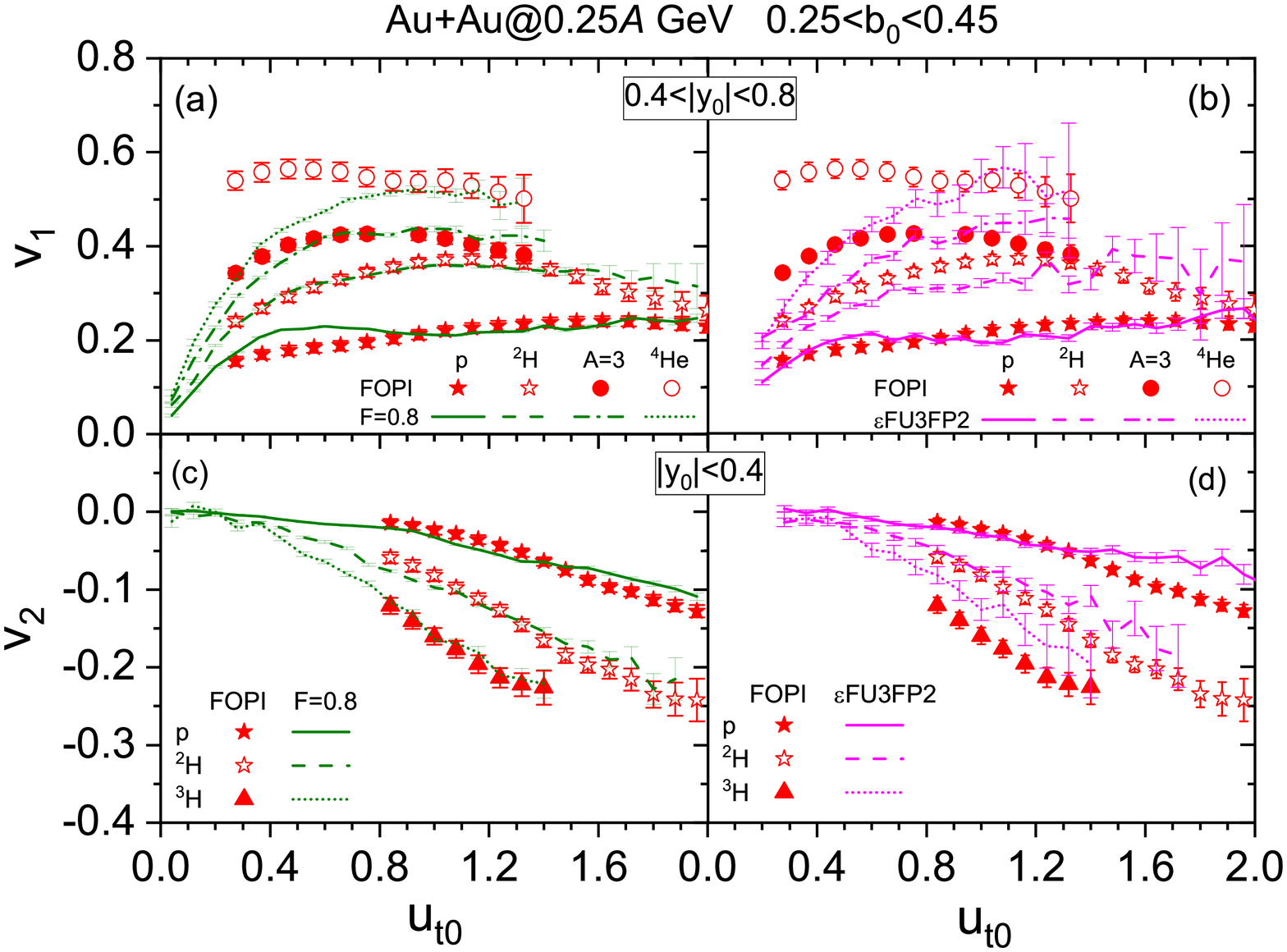}
\caption{(Color online) Panels (a) and (b): the directed flow $v_1$ for free protons, deuterons, A=3 clusters and $^{4}\text{He}$ as a function of  
transverse 4-velocities $u_{t0}$ for semi-central Au+Au collisions at $0.25A$ GeV.
Panels (c) and (d): the same as panels (a) and (b) but for the $v_2$ of free protons, deuterons, and tritons.}
\label{v12ut0}
\end{center}
\end{figure}

In order to verify the effectiveness of the fixed equivalent in-medium correction factors $F$, and of $\varepsilon$FU3FP2 in describing the FOPI experimental data \cite{FOPI:2011aa}, Fig. \ref{v12ut0} shows the $u_{t0}$ dependence of the $v_{1}$ [panels (a) and (b)] and the $v_{2}$ [panels (c) and (d)] of light charged particles in semi-central Au+Au collisions at $0.25A$ GeV as calculated with $F=0.8$ and $\varepsilon$FU3FP2.
It is observed that both $v_{1}$ and $v_{2}$ can be reproduced fairly well, including the data of light mass fragments, such as the flows of $^{2}$H and $^{3}$H particles.
However, the experimental data of $v_{1}$ of free protons and $^{4}$He particles can not be well described when $u_{t0}<0.8$, it might be due to the deficiency of $^{4}$He which is produced from heavier excited fragments and its instability after production in model simulations.
Moreover, in QMD-like model, the yield of free nucleons is commonly overestimated while intermediate mass fragments is underestimated, due to simplifications in the initial wave function of particles and quantum effects in two-body collisions \cite{Wang:2013wca}. It implies that some of the free nucleons might belong to fragments.
In addition, since the fragments flow effects are larger than that of free nucleons, the calculated free protons flows are consequently overestimated and the calculated fragments flows be underestimated.

\subsection{Nuclear stopping}

\begin{figure}[b]
\begin{center}
\includegraphics*[scale=0.35]{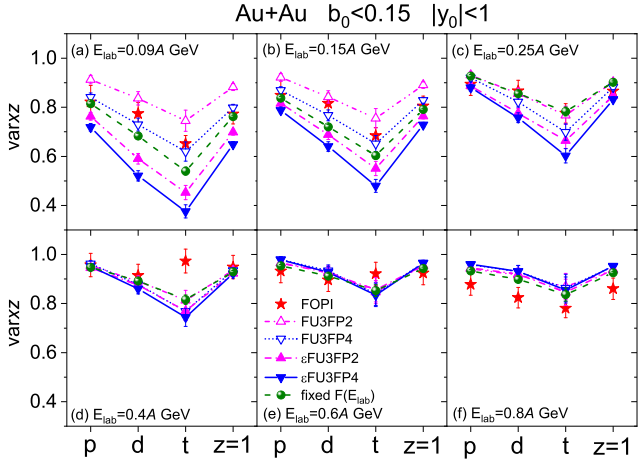}
\caption{(Color online) The nuclear stopping $varxz$ for protons, deuterons, tritons and Z=1 particles in central Au+Au collisions at the beam energies from $0.09$ to $0.8A$ GeV. Calculations with 5 in-medium correction factors indicated by different lines are compared with the FOPI experimental data (stars) \cite{FOPI:2010xrt}.
}
\label{varxzE}
\end{center}
\end{figure}

The nuclear stopping is closely related to the nucleon-nucleon collisions and in-medium effects \cite{Lopez:2014dga,Basrak:2016cbo}.
Fig. \ref{varxzE} displays the nuclear stopping observable $varxz$ of free protons, deuterons, tritons,
as well as hydrogen isotopes ($Z=1$) in central Au+Au collisions.
In addition to the FUFP and $\varepsilon$FUFP sets, the fixed equivalent and energy-dependent factors $F$, which are extracted from Fig. \ref{varxz-e}
and listed in Tab. \ref{tab1}, are taken into use as well. Again, it is found that the differences among the results calculated with $\varepsilon$FUFP and FUFP sets are well separated at lower energies but almost overlapped at higher energies. The FOPI data for $varxz$ of free protons, deuterons, tritons,
as well as hydrogen isotopes can be fairly well reproduced with the calculations using the fixed equivalent factors (shown in Tab. \ref{tab1} and Fig. \ref{dNcoll}).
It means a consistent description on both the collective flow and nuclear stopping at beam energy below  $0.8A$ GeV is achieved. At $0.8A$ GeV, $varxz$ calculated with different in-medium reduction factors on $NN$ elastic collisions are close to each other, but overestimate the nuclear stopping.

\subsection{Energy dependence of the reduction factors}\label{sec33}

\begin{figure}[t]
\begin{center}
\includegraphics*[scale=0.36]{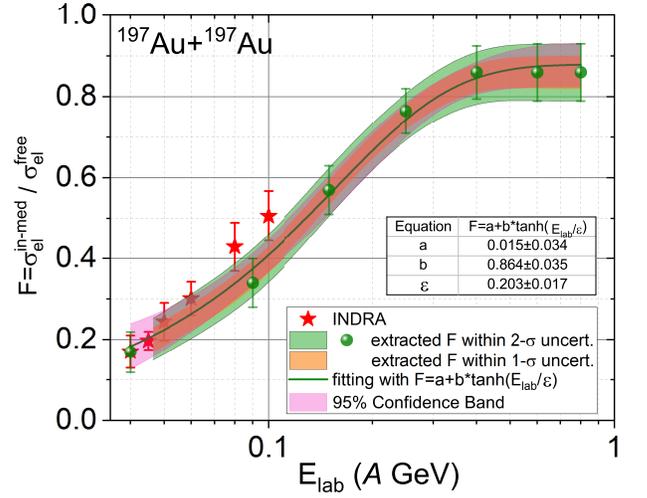}
\caption{(Color online) The excitation function of in-medium reduction factors for the $NNECS$ in nuclear matter.}
\label{factor-E}
\end{center}
\end{figure}

As shown in Fig. \ref{varxz-e}, there is a fairly well linear relationship between the collective flows and the in-medium correction factors.
By comparing the existing FOPI experimental data for proton flows \cite{FOPI:2011aa,LeFevre:2016vpp}
the UrQMD model, the fixed equivalent and energy-dependent in-medium correction factors $F$ can be constrained from $\chi^2$ analysis,
and is shown in Fig. \ref{factor-E}, represented by olive solid circles.
The red star symbols represent the results from INDRA Collaboration \cite{Lopez:2014dga}.
The olive and orange bands represent the extracted fixed $F$ within 1-$\sigma$ and 2-$\sigma$ uncertainty, respectively.
The solid line represents nonlinear fits to the extracted $F$ with 1-$\sigma$ uncertainty with assuming $F=a+b\cdot\tanh(E_{\rm{lab}}/\varepsilon)$ and the parameters can be found in Fig. \ref{factor-E}.
The pink band is 95\% confidence intervals around the fitted lines.
Let us finally mention that although the form of this energy-dependent in-medium correction factors is simple and rough, it provides intuitionistic and quantitative comprehension of in-medium effects on $NN$ elastic collisions, and can be easily incorporated in transport model.

It is inevitable that at the higher energies studied in this work the $NN$ inelastic scattering will occur, but its proportion is still not high \cite{bass1995,Liu:2020jbg}. For example, the rate for $NN\rightarrow NR$ channels (where R denotes $\Delta$ or $N^*$ resonances) is about $\sim10$\% while $\sim79$\% for $NN\rightarrow NN$ elastic collisions for Au+Au collisions at $0.8A$ GeV with $F$=1.0. Indisputably, $NN\rightarrow N\Delta$ will be influenced by nuclear medium, but the present considerations of in-medium effects on $NN\rightarrow N\Delta$ cross section at intermediate energies are different
\cite{NPA4901988,Liu:2020jbg,Yong:2017cdl,Xie:2018egs,Yuan:2010ad}. To investigate the effect of in-medium $NN$ inelastic cross section on collective flows and nuclear stopping, Au+Au collisions at $E_{lab}=0.8A$ GeV with the incompressibility $K_{0}=200$ MeV, and the correction factor $F=0.8$ for $NNECS$ while $F_{inel}=\sigma_{inel}^{in-med}/\sigma_{inel}^{free}=0.5\sim3.0$ for $NN$ inelastic cross section are performed. The same equivalent medium correction method as that for $NN$ elastic is used here for the $NN$ inelastic cross-section for simplicity.
It is found that the values of $v_{11}$ and $varxz$ increase slightly with the increase of $F_{inel}$, while $v_{20}$ keeps almost identical. This is because the enhanced $NN$ cross section leads to a greater stopping \cite{Li:2018wpv,Yong:2010zg,Liu:2001uc}. Furthermore, the enhancement of the values of $v_{11}$ of protons and nuclear stopping $varxz$ caused by the in-medium $NN$ inelastic cross section is much weaker than that caused by in-medium $NNECS$, thus the in-medium correction factor $F$ on $NNECS$, extracted from the comparison of the experimental data with the calculated collective flows of protons, will be only slightly affected by the in-medium correction on $NN$ inelastic cross section in the energy region studied in this work.

\section{Summary and outlook}\label{sec4}

In summary, this work studies the energy dependence of the in-medium correction factor of $NN$ elastic cross section, and its effects on the collective flows and nuclear stopping in $^{197}$Au+$^{197}$Au collisions at beam energies from $0.09$ to $0.8A$ GeV with using the UrQMD model.
By introducing the energy dependence into the parameterized in-medium correction factors (FUFP set) of $NNECS$, which depends on the density and the momentum, the experimental data of collective flows and nuclear stopping are compared with that from simulations with different in-medium correction factors.
It is clearly seen that the energy dependence of the in-medium correction factor of $NNECS$ has an obvious influence on collective flows and nuclear stopping in HICs at relatively low energies.

In addition, a fairly well linear relationship between the in-medium correction factor on $NNECS$ and the collective flows is found.
Then, the fixed equivalent and beam energy-dependent in-medium correction factors $F$ is extracted based on the comparison of the FOPI experimental data on the collective flows with the UrQMD model simulations.
It is more exciting that the extracted $F$ is also consistent with previous results from INDRA Collaboration by using nuclear stopping as the probe \cite{Lopez:2014dga}, the in-medium effects give a significant reduction of the $NNECS$, and this effect decreases with increasing beam energy ($\sim$80\% at $E_{lab}=0.04A$ GeV and $\sim$24\% at $0.25A$ GeV).
Finally, a phenomenological formula for in-medium $NNECS$ is presented by fitting the extracted $F$, and this formula can be easily incorporated in transport model. Additional calculations with other transport models by using the extracted $F$ will be useful for ascertaining the present results.

Although the influence of the nuclear incompressibility, Pauli-blocked algorithm and in-medium $NN$ inelastic cross section on the extracted $F$ is insignificant separately, but there might be a coupling effect to some extent among these quantities might influence the extracted $F$ value somewhere visibly and deserve further studies, even with the help of the modern machine learning technique. In addition, a more consistent treatment of the medium effects on both nucleons and $\Delta$ baryons and $\pi$ mesons in a transport model is still urgently required in order to achieve a more systematic description on HICs at SIS energies.

\section*{Acknowledgements}

The authors are grateful to the C3S2 computing center in Huzhou University for calculation support. The work is supported in part by the National Natural Science Foundation of China (Nos. 11875125, U2032145, 12175170 and 12147219), the National Key Research and Development Program of China under Grant No. 2020YFE0202002. P. Li gratefully acknowledges the financial support from China Scholarship Council (No. 202106180053) and the warm hospitality of the Institut f\"{u}r Theoretische Physik, Johann Wolfgang Goethe Universit\"{a}t, Germany.

\nolinenumbers


\end{document}